\documentclass{aa}
\usepackage{graphicx}
\usepackage{subcaption}
\usepackage[colorlinks=true, citecolor=blue, linkcolor=blue, urlcolor=black]{hyperref}
\usepackage[varg]{txfonts}
\usepackage{comment}
\usepackage{hyperref}
\usepackage{mathrsfs}
\usepackage{amsmath}

\begin{document}

   \title{Low frequency radio properties of the $z>5$ quasar population}
   \author{A. J. Gloudemans \inst{\ref{inst1}}
   \and K. J. Duncan \inst{\ref{inst2}} 
   \and H. J. A. R\"{o}ttgering \inst{\ref{inst1}}
   \and T. W. Shimwell \inst{\ref{inst1}, \ref{inst3}}
   \and B. P. Venemans \inst{\ref{inst1}}
   \and P. N. Best \inst{\ref{inst2}}
   \and \\ M. Br\"{u}ggen \inst{\ref{inst4}}
   \and G. Calistro Rivera \inst{\ref{eso}}
   \and A. Drabent \inst{\ref{inst5}}
   \and M. J. Hardcastle \inst{\ref{inst_Hertfordshire}}
   \and G. K. Miley \inst{\ref{inst1}}
   \and D. J. Schwarz \inst{\ref{inst6}}
   \and \\ A. Saxena \inst{\ref{inst7}}
   \and D. J. B. Smith \inst{\ref{inst_Hertfordshire}}
   \and W. L. Williams \inst{\ref{inst1}}}

   \institute{Leiden Observatory, Leiden University, PO Box 9513, 2300 RA Leiden, The Netherlands \\ e-mail: gloudemans@strw.leidenuniv.nl\label{inst1} 
   \and SUPA, Institute for Astronomy, Royal Observatory, Blackford Hill, Edinburgh, EH9 3HJ, UK\label{inst2}
   \and ASTRON, Netherlands Institute for Radio Astronomy, Oude Hoogeveensedijk 4, 7991 PD, Dwingeloo, The Netherlands\label{inst3}
   \and University of Hamburg, Gojenbergsweg 112, 21029 Hamburg, Germany \label{inst4}
   \and European Southern Observatory, Karl-Schwarzchild-Strasse 2, 85748, Garching bei M\"unchen, Germany \label{eso}
   \and Th\"{u}ringer Landessternwarte, Sternwarte 5, D-07778 Tautenburg, Germany \label{inst5}
   \and Centre for Astrophysics Research, University of Hertfordshire, Hatfield, AL10 9AB, UK \label{inst_Hertfordshire}
   \and Fakult\"{a}t für Physik, Universität Bielefeld, Postfach 100131, 33501 Bielefeld, Germany \label{inst6}
   \and Department of Physics and Astronomy, University College London, Gower Street, London WC1E 6BT, UK \label{inst7}}
   
   \date{Received 6 July 2021 / Accepted 7 October 2021}
   
 \abstract{Optically luminous quasars at $z > 5$ are important probes of super-massive black hole (SMBH) formation. With new and future radio facilities, the discovery of the brightest low-frequency radio sources in this epoch would be an important new probe of cosmic reionization through 21-cm absorption experiments. In this work, we systematically study the low-frequency radio properties of a sample of 115 known spectroscopically confirmed $z>5$ quasars using the second data release of the Low Frequency Array (LOFAR) Two Metre Sky survey (LoTSS-DR2), reaching noise levels of $\sim$80 $\mu$Jy beam$^{-1}$ (at 144 MHz) over an area of $\sim5720$ deg$^2$. We find that 41 sources (36\%) are detected in LoTSS-DR2 at $>2 \sigma$ significance and we explore the evolution of their radio properties (power, spectral index, and radio loudness) as a function of redshift and rest-frame ultra-violet properties. We obtain a median spectral index of $-0.29^{+0.10}_{-0.09}$ by stacking 93 quasars using LoTSS-DR2 and Faint Images of the Radio Sky at Twenty Centimetres (FIRST) data at 1.4 GHz, in line with observations of quasars at $z<3$. We compare the radio loudness of the high-$z$ quasar sample to a lower-$z$ quasar sample at $z\sim2$ and find that the two radio loudness distributions are consistent with no evolution, although the low number of high-z quasars means that we cannot rule out weak evolution. Furthermore, we make a first order empirical estimate of the $z=6$ quasar radio luminosity function, which is used to derive the expected number of high-$z$ sources that will be detected in the completed LoTSS survey. This work highlights the fact that new deep radio observations can be a valuable tool in selecting high-$z$ quasar candidates for follow-up spectroscopic observations by decreasing contamination of stellar dwarfs and reducing possible selection biases introduced by strict colour cuts.}


\maketitle

\section{Introduction}
\label{sec:introduction}

High redshift quasars are important tools for studying black hole formation and evolution and the process of cosmic reionization (e.g. \citealt{Fan2006ARA&A..44..415F, Volonteri2012Sci...337..544V, schroeder2013evidence, Venemans2015ApJ...801L..11V, mortlock2016ASSL..423..187M}). 
Through Ly$\alpha$ forest and Gunn-Peterson trough measurements \citep{Gunn1965ApJ...142.1633G}, cosmic reionization has been strongly constrained to end at $z\approx6$ (e.g. \citealt{Becker2001AJ....122.2850B, Fan2001AJ....122.2833F, Fan2006ARA&A..44..415F, McGreer2013, schroeder2013evidence}). However, the saturation of Ly$\alpha$ absorption at neutral fractions above $x_{\text{HI}}\sim$10$^{-4}$ means that detailed constraints on the earlier stages of reionization become more difficult (see e.g. \citealt{Santos2004MNRAS.349.1137S,Fan2006ARA&A..44..415F, Maselli2007MNRAS.376L..34M, Mesinger2008MNRAS.386.1990M}), although at neutral fractions $x_{\text{HI}}>$0.1 the Ly$\alpha$ damping wing seen in quasar spectra has been successfully used to provide constraints on the neutral fraction up to $z=7.5$ (see e.g. \citealt{Mortlock2011Natur.474..616M,Greig2017MNRAS.466.4239G, Banados2018Natur.553..473B}). To be able to directly probe the neutral hydrogen in the Epoch of Reionization (EoR), it has been suggested that luminous radio sources, such as the most radio-luminous quasars, can be used to measure the column density of neutral hydrogen via 21-cm absorption studies \citep{carilli2002ApJ...577...22C, Mack2012MNRAS.425.2988M}. The 21-cm absorption line does not saturate and could therefore be used to provide direct measurements of the reionisation process as a function of redshift. 

Traditionally, quasars have been divided into classes of radio-loud (RL) and radio-quiet (RQ) depending on the ratio of their radio and optical flux densities and luminosities (e.g. \citealt{Kellermann1964ApJ...140..969K, Strittmatter1980A&A....88L..12S, Kellermann1989AJ.....98.1195K}). It is, however, unclear whether these two classes are probing a different population of quasars or if any bimodality seen is actually caused by an underlying asymmetric radio luminosity distribution and observational biases induced by flux-limited samples (e.g. \citealt{Ivezic2002AJ....124.2364I, Kimball2011ApJ...739L..29K, Balokovic2012ApJ...759...30B, gurkan2019A&A...622A..11G, Ceraj2020A&A...642A.125C, Macfarlane2021}). Depending on the sample redshift and optical (or UV) luminosity, the RL fractions measured in the literature vary from 4.1 to 24.3\% \citep{Padovani1993MNRAS.263..461P, Jiang2007ApJ...656..680J, Balokovic2012ApJ...759...30B, Banados2015ApJ...804..118B}, with interpretation of the observed variations made more complex due to the range of definitions for the RL fraction and different wavelengths used. Furthermore, as radio observations have increased in sensitivity, RQ quasars have also been shown to exhibit radio emission (e.g. \citealt{Kellermann1989AJ.....98.1195K, Doi2013ApJ...765...69D}) that could be produced by either star formation or weak radio jets (e.g. \citealt{Kimball2011ApJ...739L..29K, Zakamska2016MNRAS.455.4191Z, Fawcett2020MNRAS.494.4802F, Rosario2020MNRAS.494.3061R, Macfarlane2021}).

\cite{gurkan2019A&A...622A..11G} and \cite{Macfarlane2021} have investigated the dependence of the radio loudness of quasars on their properties, such as redshift, black hole mass, and optical and radio luminosity. \citet{Macfarlane2021} suggest that both AGN jets and star formation contribute to the observed radio emission with a smooth transition from the jet to star formation dominated regime. However, many outstanding questions in the field remain, such as what physical mechanisms are involved in the generation of radio jets in quasars, how these evolve as a function of redshift and SMBH properties, and why powerful radio jets are only prevalent in a small fraction of the quasar population. Observationally constraining the distribution of radio luminosities at high redshift is therefore essential for both understanding the physical mechanisms behind radio jets and accurately quantifying the prevalence of bright radio sources suitable for 21-cm cosmology experiments.

Recently, \cite{Belladitta2020A&A...635L...7B} discovered the brightest radio-loud AGN at $z>6$ (with a flux density of 23.7 mJy at 1.4 GHz) and studies of the properties of the high-$z$ ($z>5.5$) radio-loud quasar population have been conducted by for example \cite{Hook1998ASPC..146...17H} and \cite{Banados2015ApJ...804..118B}. However, these studies have been carried out using high-frequency radio observations (>1.4 GHz) and there is only limited information on the radio properties of quasars in the low-frequency regime ($<200$ MHz) that probes rest-frame 21-cm at the redshifts of the EoR. The forthcoming second data release of the LOFAR Two Metre Sky survey (LoTSS-DR2; Shimwell et al. in prep) is currently the largest and most sensitive survey in this regime, reaching typical noise levels of $\sim$83 $\mu$Jy beam$^{-1}$ at 120--168 MHz over a wide-field area of $\sim$5700 deg$^2$. The sensitivity and coverage offered by LoTSS-DR2 allows us to systematically study rare radio quasars with a statistically significant number of sources, which was not possible before. In this work, we characterise the low-frequency radio properties of the known $z>5$ quasar population.

This paper is structured as follows. The high-$z$ quasar sample and the data used from radio and optical/infrared surveys are outlined in Sect.~\ref{sec:data}. We explore the low-frequency properties of 115 spectroscopically confirmed quasars at $z>5$ within the LoTSS-DR2 footprint in Sect.~\ref{sec:properties_quasars_lotss}, including a measurement of the median radio spectral index obtained by stacking. 
In Sect.~\ref{sec:radio_loudness} we investigate the radio loudness distribution of the quasar sample and its evolution with redshift. Based on the radio luminosity function models at lower-$z$ of \cite{Macfarlane2021} and the UV luminosity function of \cite{Matsuoka2018ApJ...869..150M}, we then make predictions of the radio luminosity function expected for high-$z$ quasars (Sect.~\ref{sec:predictions}). 
Finally, our summary is presented in Sect.~\ref{sec:conclusions}. Throughout this work, we use the AB magnitude system \citep{Oke1983ApJ...266..713O} and assume a $\Lambda$-CDM cosmology using H$_{0}$= 70 km s$^{-1}$ Mpc$^{-1}$, $\Omega_{M}$ = 0.3, and $\Omega_{\Lambda}$ = 0.7. 

\section{Data}
\label{sec:data}

\subsection{Quasar sample}
\label{subsec:quasar_sample}

\begin{figure}
\centering
   \includegraphics[width=\columnwidth, trim={0.0cm 0cm 0cm 0.0cm}, clip]{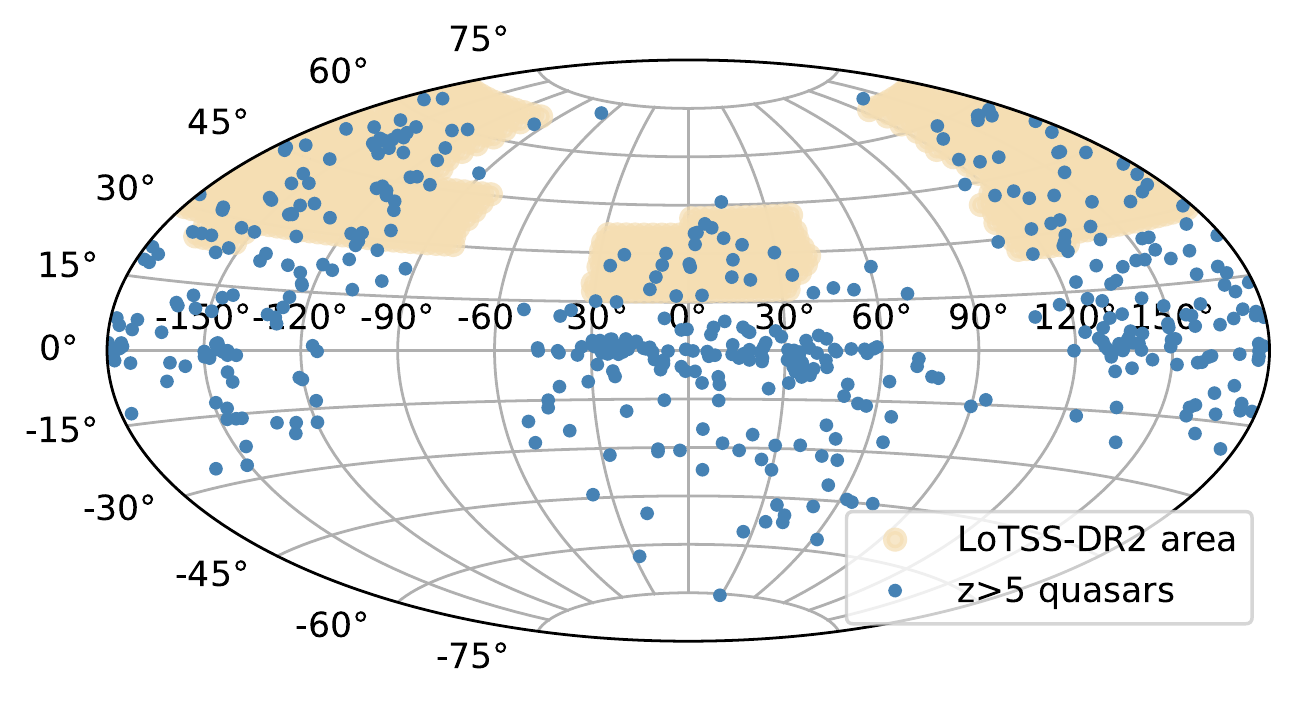}
     \caption{Sky positions of known quasars in the $z>5$ sample of \citet[][blue circles]{Ross2020MNRAS.494..789R}. The LoTSS-DR2 area is illustrated as the beige shaded and encompasses $\sim$5720 deg$^2$ of the Northern sky, split across both the North and South Galactic Caps.} 
     \label{fig:quasars_sky}
\end{figure}

In this work, we use the compiled high-$z$ quasar sample of \cite{Ross2020MNRAS.494..789R}. Their publicly available catalogue\footnote{\url{www.github.com/d80b2t/VHzQ}} currently consists of 488 spectroscopic confirmed quasars at $z>5$. These quasars were discovered in 21 surveys (ATLAS; \citealt{Shanks2015}, CFHQS; \citealt{Willott2007}, DELS; \citealt{Dey2018}, ELAIS; \citealt{Vaisanen2000}, FIRST; \citealt{Becker1995}, HSC; \citealt{Miyazaki2018}, IMS; \citealt{Kim2015}, MMT; \citealt{McGreer2013}, NDWFS; \citealt{Jannuzi_Dey1999}, PSO; \citealt{Kaiser2002, Kaiser2010}, RD; \citealt{Mahabal2005}, SDSS; \citealt{stoughton2002AJ....123..485S}, SDWISE; \citealt{WangF2016},  SHELLQs; \citealt{Matsuoka2016ApJ...828...26M}, SUV; \citealt{YangJ2017}, UHS; \citealt{WangF2017}, ULAS; \citealt{Lawrence2007}, VDES; \citealt{Reed2017}, VHS; \citealt{McMahon2013Msngr.154...35M}, VIK; \citealt{Edge2013}, VIMOS; \citealt{LeFevre2003}). These surveys cover a wide range of wavelengths and the quasars have been selected using optical, infrared, and radio observations. The redshift of all quasars have been confirmed using optical and near-infrared spectroscopy, which has also been used to determine the positions of the quasars. The sky positions of these quasars and the LoTSS-DR2 area, further discussed in Sect.~\ref{subsubsection:lofar}, are shown in Fig.~\ref{fig:quasars_sky}. 
While more complete samples of the known high-$z$ quasar population exist (e.g. \citealt{sarah_e_i_bosman_2021_4553781}), the compilation from \cite{Ross2020MNRAS.494..789R} benefits from rigorous checking of the coordinates and duplications. Furthermore, 
\cite{Ross2020MNRAS.494..789R} studied the near-infrared (NIR) and mid-infrared (MIR) properties of these known quasars and compiled a consistent and homogeneous catalogue, which is also publicly available and forms the basis of key parts of our subsequent analysis (see Sect.~\ref{subsec:optical_ir_data}). We refer the reader to \cite{Ross2020MNRAS.494..789R} for additional details on the production of this catalogue.

\subsection{Optical and infrared data}
\label{subsec:optical_ir_data}

While this paper primarily concerns the low-frequency radio properties of the $z > 5$ quasar sample, it is important to relate these radio measurements to the rest-frame optical to UV properties typically used to quantify and study the high-$z$ quasar population. To investigate observational and physical properties of the quasars we exploit the additional photometry generated by \cite{Ross2020MNRAS.494..789R}. In the northern hemisphere, this catalogue contains photometric measurements of the $Y$, $J$ and $K_{s}$ bands from the UK Infra-Red Telescope (UKIRT) Infrared Deep Sky Survey (UKIDSS; \citealt{Lawrence2007MNRAS.379.1599L}).
For the majority of sources in this analysis, the $J$-band photometry is provided by the UKIDSS Hemisphere Survey (UHS; \citealt{Dye2018MNRAS.473.5113D}).
In addition to the NIR observations, \citet{Ross2020MNRAS.494..789R} also compiled MIR observations from the Wide-field Infrared Survey Explorer (WISE; Wright et al. 2010). 
The WISE observations are taken from the unWISE \citep[][3.4 and 4.6 $\mu$m]{schlafly2019ApJS..240...30S} and AllWISE data catalogues \citep[][12 and 23 $\mu$m]{Cutri2014yCat.2328....0C}. 

To supplement the NIR and MIR observations, we cross-match the quasar sample with the deepest available large area optical surveys available within the LoTSS-DR2 footprint. These consist firstly of the DESI Legacy Imaging Surveys \citep[][specifically Data Release 8]{dey2019AJ....157..168D}, which provides $z$-band photometry reaching a 5$\sigma$ average depth of $\sim$22.5 mag but are limited to low extinction areas, and secondly, the Panoramic Survey Telescope and Rapid Response System (PANSTARRS 1; \citealt{Chambers2016arXiv161205560C}) survey which covers the full northern hemisphere and provides $z$ and $y$-band measurements down to typical 5$\sigma$ depths of 20.9 and 19.7 AB mag, respectively. An additional benefit of the Legacy data is that for all sources with optical detections, it provides forced deblended WISE photometry, offering more reliable MIR measurements. We use the deblended WISE measurements if available, and otherwise use the measurements provided in \cite{Ross2020MNRAS.494..789R}. All the quasars in our sample are unresolved so the various photometric measures described above provide estimates of their total fluxes in each band.

\subsection{Radio data}
\label{subsec:lofar_obs}

\subsubsection{LOFAR}
\label{subsubsection:lofar}

LoTSS is an ongoing LOFAR survey at 120-168 MHz (144 MHz central frequency) with the aim to eventually observe the whole Northern sky with a 6$\arcsec$ angular resolution and $\sim$100 $\mu$Jy beam$^{-1}$ RMS. The previous data release, DR1 (\citealt{Shimwell2019A&A...622A...1S}), covered 424 deg$^2$, whereas the forthcoming DR2 spans an area of 5720 deg$^2$ of the Northern sky with a total of $\sim4.4$ million detected radio sources (Shimwell et al. in prep). The LoTSS-DR2 data release consists of two fields denoted the 0h and 13h fields (see Fig.~\ref{fig:quasars_sky}) and reaches a median sensitivity of $\sim$83 $\mu$Jy beam$^{-1}$. Compared to the Faint Images of the Radio Sky at Twenty Centimetres (FIRST) survey \citep{Becker1994ASPC...61..165B} conducted with the Karl G. Jansky Very Large Array (VLA), the LoTSS survey is about 10 times more sensitive for a compact radio source with a spectral slope of $\alpha=-0.7$, where $S_{\nu} \propto \nu^{\alpha}$. 

Of our full sample of 488 quasars, 115 lie within the LoTSS-DR2 footprint. The LoTSS-DR2 source catalogue contains all radio detections at $>5\sigma$ above the local noise background, which has a median value of 74 $\mu$Jy and where local is defined as a box of $\sim$3$\arcmin$x3$\arcmin$ (Shimwell et al. in prep). Of our sample, we find a total of 19 detected with this significance. However, to also investigate the faint radio emission from quasars below this limit, we perform forced photometry of the radio flux density by taking the peak pixel value at each quasar location in the sample. 
The errors on the peak flux density are estimated by measuring the standard deviation using the median absolute deviation within $\sim$150$\arcsec$ of the source location.
We note that both the \cite{Ross2020MNRAS.494..789R} and LoTSS-DR2 source positions are tied to the Gaia reference frame \citep{Gaia2016A&A...595A...2G}, and the astrometric uncertainty in LoTSS-DR2 of $\lesssim$0.2$\arcsec$ is significantly below the LoTSS pixel scale of 1.5$\arcsec$.
We can therefore be confident that the peak flux measurements are not significantly impacted by systematic or statistical astrometric offsets between the optical and radio datasets.

Based on the peak flux density measurements at the known quasar positions and assuming unresolved sources, we find a total of 41 tentative detections at $>2 \sigma$ significance (including the 19 $>5\sigma$ detected), where the errors are based on the local RMS. The detection fraction of known $z>5$ quasars in LoTSS-DR2 at $>2\sigma$ is therefore 36\%. 
In this work, the peak flux densities are corrected to total flux density measurements based on the average ratio of forced peak flux density and total flux density (factor $\sim$1.4) for the 19 $>5\sigma$ detected quasars in LoTSS-DR2. This factor is similar to the peak to total flux density ratio found for all unresolved sources in the LoTSS DR2 catalog of $\sim$1.3. 
While theoretically the peak and total flux density should be equal for unresolved sources, due to issues in the calibration the sources are not fully focused. Finally, to account for any systematic offset in flux calibration, a flux density scale uncertainty of 10\% is added to the error in quadrature when converting to luminosity.

We note that additional LOFAR observations do exist for many fields, the deepest of which are in the LoTSS deep fields ELAIS-N1, Lockman Hole, and Bo\"otes reaching $\sim$25 $\mu$Jy/beam noise level \citep{Tasse2021A&A...648A...1T, Sabater2021A&A...648A...2S}, so the potential LOFAR detected quasar sample could be grown. However, the areas of the both the additional LoTSS and GAMA fields total only $\sim$250 deg$^2$ and would yield only a fractional increase compared to the LoTSS-DR2 region used here. Therefore, to ensure homogeneous processing we restrict the analysis here to the LoTSS-DR2 area. 

\subsubsection{FIRST}
\label{subsec:first}
To obtain a measure of the radio spectral index (see Sect.~\ref{sec:properties_quasars_lotss}), we compare the LOFAR radio fluxes at 144 MHz to the fluxes from the FIRST survey by \cite{Becker1995} at 1.4 GHz. The FIRST survey covers over 10,000 deg$^2$ of the Northern Galactic Cap using the NROA Very Large Array at a resolution of 5$\arcsec$ with a typical rms of 0.15 mJy \citep{Becker1995}. Out of the 115 quasars, 93 are located in the FIRST area and 10 sources are detected at $>2\sigma$ significance, again using the peak pixel for the flux density measurement and the local RMS (of around 150 $\mu$Jy) for the error estimate. Similarly, out of the 41 quasars detected above $2\sigma$ by LOFAR, 35 are located in the FIRST area and 9 sources are detected at $>2\sigma$ significance in the FIRST survey. This data is used to obtain spectral indices in Sect.~\ref{sec:properties_quasars_lotss}.

\subsection{SED fitting procedure}
\label{subsec:sedfitting}

The optical to MIR observations we compile are used to provide consistent estimates of the rest-frame UV and optical magnitudes for our sample.
Typically, literature estimates of rest-frame UV magnitude are determined using the measured flux in the observed filter which most closely probes the rest-frame wavelength in question (and assuming a standard $k$-correction).
To exploit the full range of data available for our sample, we instead derive rest-frame UV to optical magnitudes by performing spectral energy distribution (SED) fitting using the template fitting code \textsc{Eazy}  \citep{brammer2011ApJ...739...24B}. This SED fitting technique is consistent across the multiple datasets within the sample and allows for variation in the UV-optical spectral slopes in individual sources. The main advantages of using this method are therefore being able to account for the individual $k$-corrections and UV power law slopes required for each source. Additionally, our analysis homogenises the measurements for quasars presented in different surveys that may have previously quoted UV-magnitudes based on a range of different datasets and photometry methods.

To derive the rest-frame UV magnitudes, we fix the redshift to the known spectroscopic redshifts and fit the observed photometry using the templates generated by \cite{Brown2019MNRAS.489.3351B}, which are derived from AGN SEDs and span a broad range of quasar types and line properties, including additional reddening to allow for the full range of spectral slopes. To ensure reliable fits, the fitting was only performed on quasars which had a detection (S/N $> 3\sigma$) in either the Legacy $z$ or UKIDSS $Y$-band, and a $>3\sigma$ detection in both the $J$ and W1-bands. From these fits we derive the rest-frame UV absolute magnitudes at wavelengths of 1450, 2500, and 4400\,$\r{A}$, which are the primary rest-UV measurements used for high-$z$ luminosity functions and the rest-frame properties commonly used to quantify the radio loudness distribution of quasars (e.g. \citealt{Banados2015ApJ...804..118B, Inayoshi2020ARA&A..58...27I}). We define these filters as tophat functions centred on the specific central wavelength with a width of 100 $\AA$. 
The rest-frame magnitudes and their associated uncertainties are obtained through a simple Monte Carlo approach.
Each source is duplicated 500 times, perturbing the fluxes in each band by drawing from a normal distribution with width determined by the measured flux errors from the catalogues. The median rest-frame magnitudes are then given by the 50th percentile value and the errors by the 16th and 84th percentile. 
We reiterate that these measurements represent the observed rest-frame magnitudes, and that intrinsic UV magnitudes will naturally be brighter due to dust reddening (e.g. \citealt{Calistro2021A&A...649A.102C}). 

\begin{figure}
\centering
   \includegraphics[width=\columnwidth, trim={0.0cm 0cm 0cm 0.0cm}, clip]{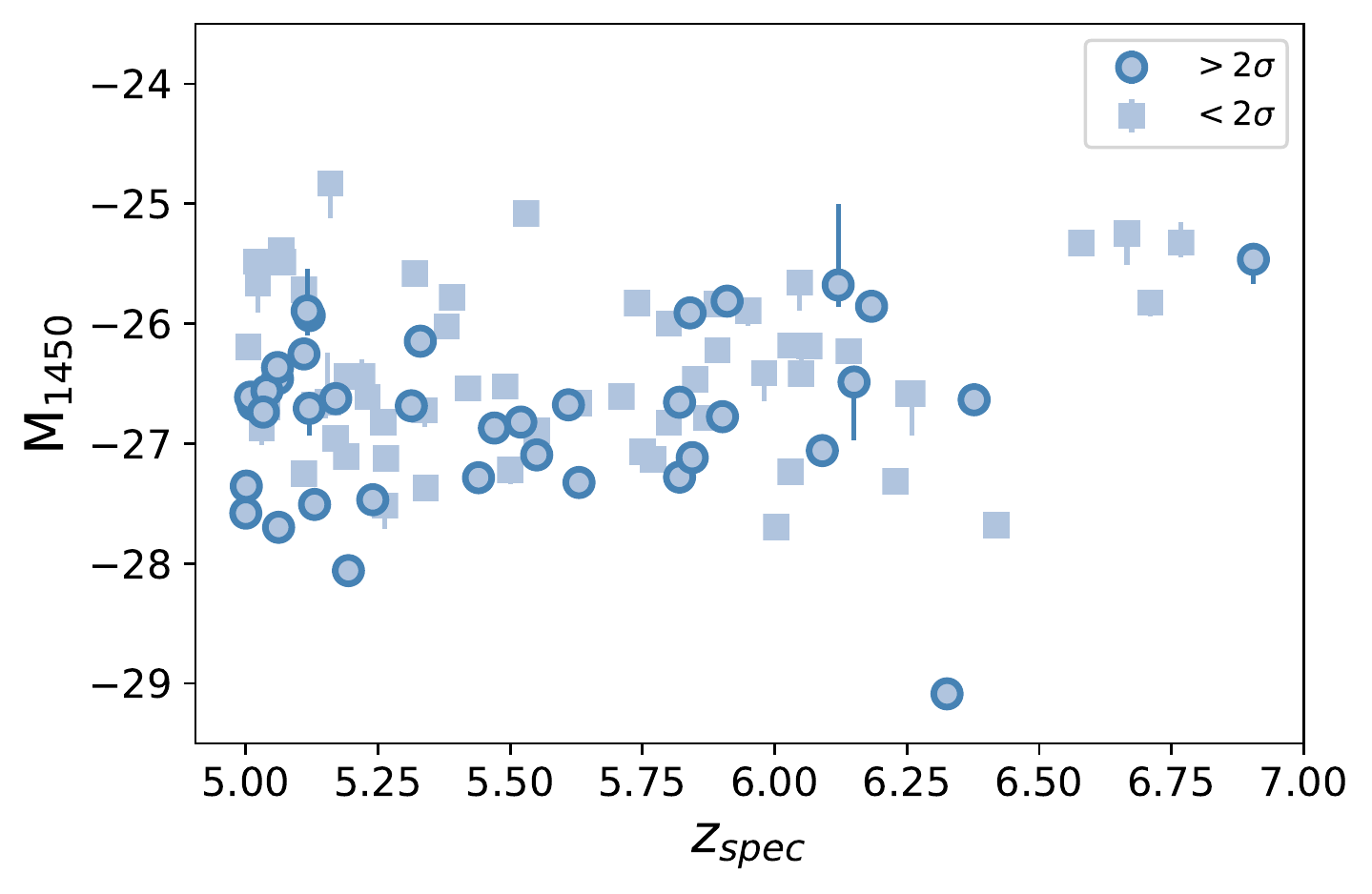}
     \caption{Rest-frame absolute UV magnitude $M_{1450}$ as a function of redshift for the 97 high-$z$ quasars in the LoTSS-DR2 area that satisfy the SED fitting criteria (see Sect.~\ref{subsec:sedfitting}). The quasars detected with > 2$\sigma$ by LOFAR are indicated by circles and the non-detected quasars by squares.}
     \label{fig:M1450_Z}
\end{figure}

Out of the 115 sources in the LoTSS-DR2 area, 97 satisfy the SED fitting criteria. Fig.~\ref{fig:M1450_Z} shows the rest-frame absolute UV magnitude at 1450$\AA$ as a function of the spectroscopic redshift for these sources. There is no evidence for any strong trends of $M_{1450}$ with redshift in this sample.

From the absolute UV magnitude ($M_{1450}$), the SMBH mass (M$_{\text{BH}}$) of the quasars can be estimated using an empirical relation derived by \cite{Inayoshi2020ARA&A..58...27I}, which assumes a constant bolometric correction and Eddington ratio and is given by
\begin{equation}
\label{eq:bhmass}
    \log_{10}\left ( \frac{\text{M}_{\text{BH}}}{\text{M}_{\odot}} \right ) = \frac{-M_{1450} - 3.46}{2.5}.
\end{equation}

Although the remainder of this work focuses only on the sources within the radio data footprint, the measured rest-frame magnitudes at 1450, 2500, and 4400 $\AA$ of the full $z > 5$ quasar sample are provided as a data product \footnote{\label{cdstable}The measured rest-frame magnitudes and LoTSS-DR2 fluxes will be available in electronic form
at the CDS via anonymous ftp to cdsarc.u-strasbg.fr (130.79.128.5)
or via \url{http://cdsweb.u-strasbg.fr/cgi-bin/qcat?J/A+A/}. For now the table can be found here \url{https://github.com/AnniekGloudemans/LOFAR_high_z_quasars}}. An example of the radio fluxes and rest-frame UV magnitudes of the first 10 radio detected high-$z$ quasars is given in Tab.~\ref{tab:quasars}.

\begin{table*}
\caption{Measured radio fluxes and rest-frame UV magnitudes of the first 10 radio detected high-$z$ quasars from the full sample of \cite{Ross2020MNRAS.494..789R}. The total fluxes are based on the assumption that the quasars are unresolved in LOFAR and have been obtained by a peak to total flux density correction of a factor 1.4 (see Sect.~\ref{subsubsection:lofar}). See footnote 2. } 
\label{tab:quasars}
\centering
\resizebox{\textwidth}{!}{
\begin{tabular}{c c c c c c c c c c}  
\hline\hline       
Survey & Name & RA (hms) & Dec (dms) & $z$ & Total flux & Total flux err & $M_{1450}$ & M$_{2500}$ & M$_{4400}$ \\ 
 &  & (J2000) & (J2000)  &  &(mJy) &  (mJy) &  & & \\ 
\hline \hline

SDSS  &  J0002+2550  &  00:02:39.39  &  +25:50:34.80  &  5.82  &  1.17  &  0.14  &  -27.28  &  -27.20  &  -27.58\\
SDSS  &  J0012+3632  &  00:12:32.88  &  +36:32:16.10  &  5.44  &  1.23  &  0.21  &  -27.28  &  -27.74  &  -27.68\\
SDSS  &  J0100+2802  &  01:00:13.03  &  +28:02:25.84  &  6.33  &  0.88  &  0.13  &  -29.09  &  -29.48  &  -29.60\\
SDSS  &  J0157+3001  &  01:57:45.45  &  +30:01:10.68  &  5.63  &  0.43  &  0.18  &  -27.32  &  -27.90  &  -28.29\\
SDSS  &  J0741+2520  &  07:41:54.72  &  +25:20:29.6  &  5.19  &  2.63  &  0.29  &  -28.06  &  -28.38  &  -28.95\\
SDSS  &  J0756+4104  &  07:56:18.13  &  +41:04:08.6  &  5.06  &  0.42  &  0.13  &  -26.46  &  -26.88  &  -26.86\\
DELS  &  J0803+3138  &  08:03:05.42  &  +31:38:34.2  &  6.38  &  1.32  &  0.12  &  -26.63  &  -26.54  &  -26.72\\
SDSS  &  J0810+5105  &  08:10:54.32  &  +51:05:40.10  &  5.82  &  0.40  &  0.14  &  -26.66  &  -26.66  &  -26.77\\
DELS  &  J0839+3900  &  08:39:46.88  &  +39:00:11.5  &  6.90  &  0.35  &  0.18  &  -25.46  &  -26.59  &  -27.82\\
SDSS  &  J0840+5624  &  08:40:35.09  &  +56:24:19.90  &  5.84  &  0.31  &  0.10  &  -27.12  &  -27.10  &  -26.82\\

\hline \hline
\end{tabular}}

\end{table*}

\section{Radio properties of high-$z$ quasars in LoTSS}
\label{sec:properties_quasars_lotss}

\begin{figure}
\centering
   \includegraphics[width=\columnwidth, trim={0.0cm 0cm 0cm 0.0cm}, clip]{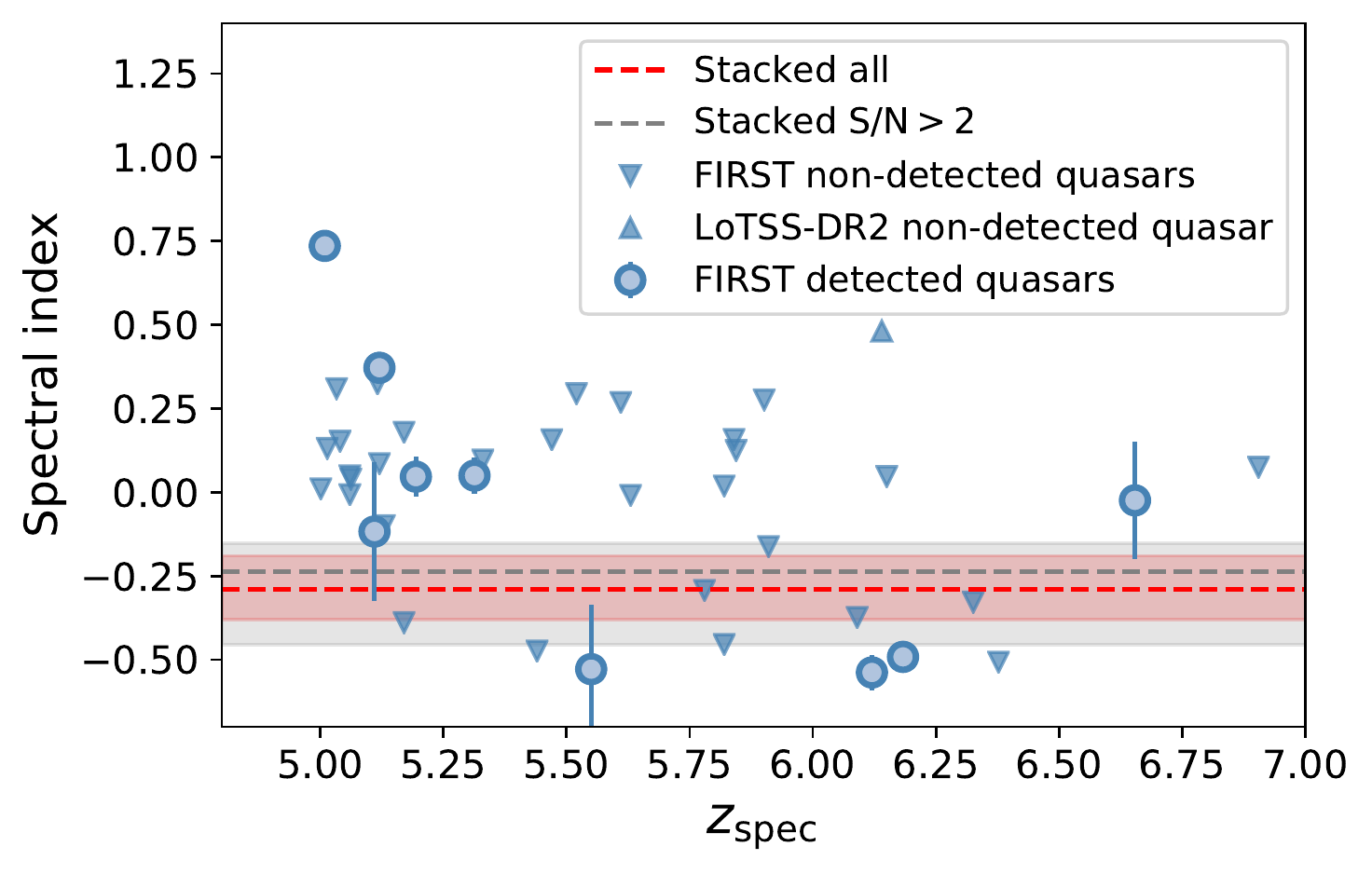}
     \caption{Radio spectral index, $\alpha$ (where $S_{\nu} \propto \nu^{\alpha}$), versus spectroscopic redshift for the 35 LOFAR detected quasars overlapping with FIRST. Upper limits are given for the quasars not detected in FIRST and a lower limit is given for the quasar not detected in LoTSS. The median spectral index for the quasar sample is both determined by stacking all 93 quasars (red) and only the 35 S/N > 2 detected ones (grey) in the overlapping LoTSS and FIRST area. The red and grey areas give the errors on the median spectral indices, which are the 16th and 84th percentiles determined by bootstrapping.} 
     \label{fig:redshift_vs_alpha}
\end{figure}

\begin{figure}
\centering
   \includegraphics[width=\columnwidth, trim={0.0cm 0cm 0cm 0.0cm}, clip]{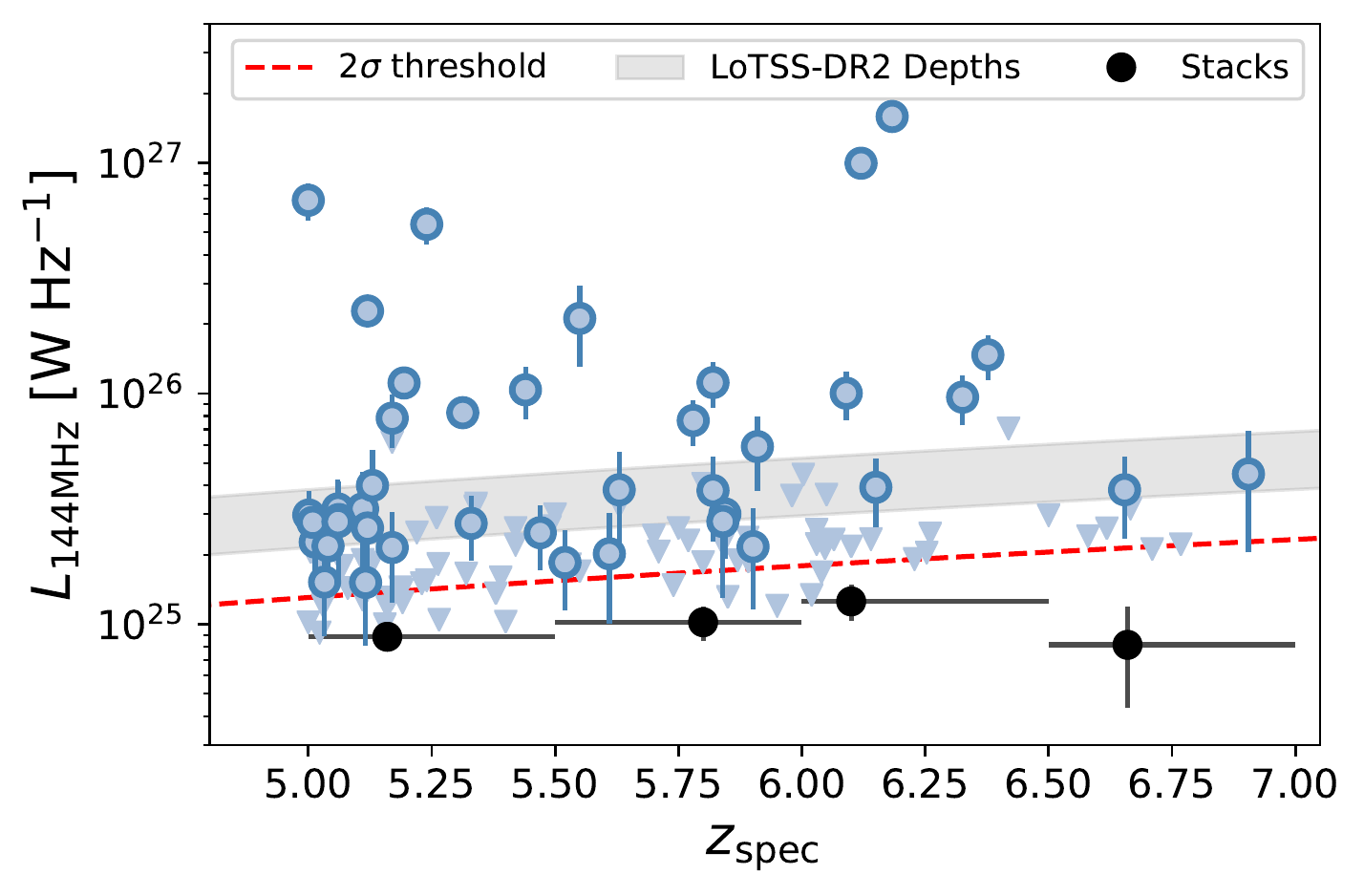}
     \caption{Radio luminosity of high-$z$ quasars at 144 MHz as a function of their redshift. The grey band indicates the $5\sigma$ detection limit range of LoTSS-DR2 (16th to 84th percentile) and the filled blue circles are quasar detections at a $>2\sigma$ level. The triangles indicate 2$\sigma$ upper limits for quasars with $<2\sigma$ detection. The black circles show the median stacked radio luminosity when stacking all quasars in redshift bins from $z=5$ to $7.0$ in steps $0.5$. The average 2$\sigma$ threshold is given by the red dashed line.} 
     \label{fig:radiolum_vs_z}
\end{figure}

\begin{figure*}
\centering
   \includegraphics[width=\textwidth, trim={0.0cm 0cm 0cm 0.0cm}, clip]{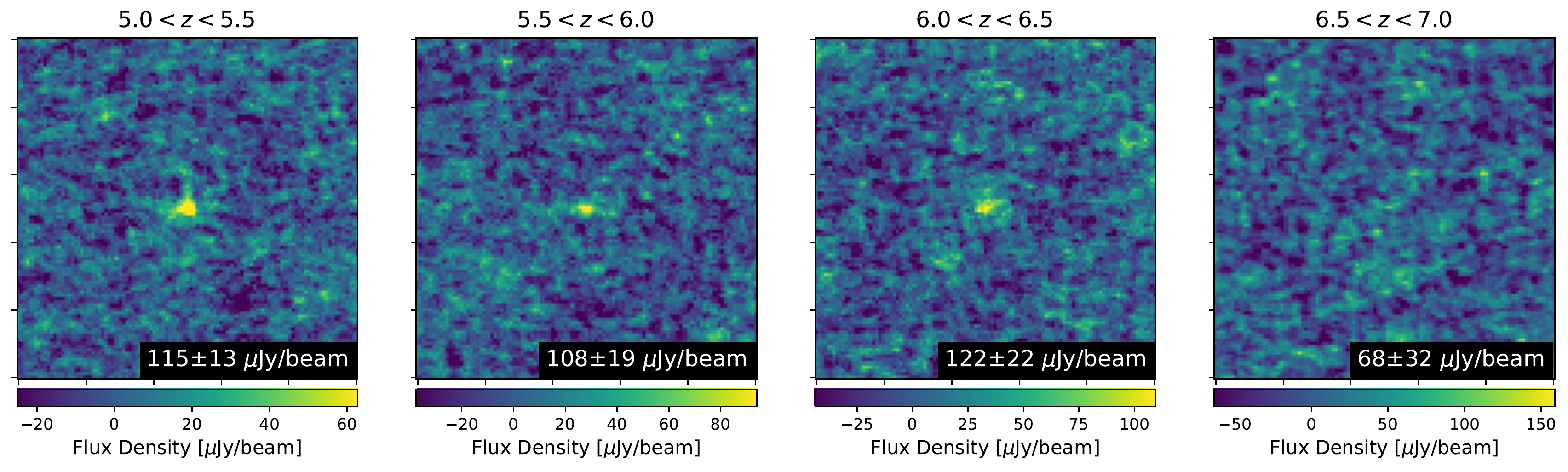}
     \caption{Median stacked LOFAR cutouts of the known quasars divided into four redshift bins. The median peak flux density and error for each stack is given in the bottom right. There is a significant radio detection ($>5\sigma$) in the three lowest redshift bins, but there is no significant detection in the $6.5<z<7$ bin.} 
     \label{fig:radiostacked_zbins}
\end{figure*}

The combination of the wide area radio surveys of FIRST at 1.4 GHz \citep{Becker1995} and LoTSS-DR2 at 144 MHz allow us to obtain a statistical measurement of the radio spectral index $\alpha$ for high-$z$ quasars at a rest-frame frequency of 5 GHz for the first time. The resulting spectral indices as a function of their spectroscopic redshift are shown in Fig.~\ref{fig:redshift_vs_alpha}. Upper limits are given for the non-detected quasars in the FIRST area, which are calculated using the 2$\sigma$ flux density limit of FIRST of 0.3 mJy. A lower limit on the spectral index ($\alpha>0.48$) of the quasar ULAS J1609+3041 is also given in Fig.~\ref{fig:redshift_vs_alpha}, which is detected in FIRST (at $\sim$3$\sigma$ significance), but not detected in LoTSS-DR2. 

To determine the median spectral index and the variation within the sample we perform a bootstrap analysis, randomly drawing from our sample (with replacement) and median stacking both the LoTSS and FIRST cutouts independently to obtain a median peak pixel flux density measurement in each survey. This is then repeated 10,000 times to produce a distribution that also incorporates uncertainties due to sample variation. The 16th and 84th percentile of the resulting peak flux density distributions, together with the systematic flux scale density uncertainty of 5\% \citep{White1997ApJ...475..479W} and 10\% (Shimwell et al. in prep) for the FIRST and LoTSS survey, respectively, are used to determine the error on the spectral index. This stacking procedure is done for all 93 quasars that overlap with FIRST (red) and for only the 35 sources detected by LoTSS with S/N > 2 that overlap with FIRST (grey), resulting in spectral indices of $-0.29^{+0.10}_{-0.09}$ and $-0.24^{+0.09}_{-0.22}$, respectively. Six of the FIRST detected quasars have spectral indices higher than these stacked values, which is as expected because these flat or positive spectral index quasars are naturally selected because of the FIRST flux limit of $\sim$144 $\mu$Jy, which is shallower than the LOFAR flux limit assuming a negative spectral index. In general, LOFAR detected quasars are brighter at 144 MHz than the non-detected ones, therefore one might expect the stacked S/N>2 sample to result in a steeper spectral index than when stacking all quasars. However, given the uncertainties on the median spectral index there is no significant deviation found between the two. In this work, we therefore assume a spectral index of $-0.29\pm$0.10 for the quasars not detected by FIRST when converting radio fluxes to luminosities. This spectral index is in line with the previous work of \cite{gurkan2019A&A...622A..11G}, who found a median spectral index of $-0.26\pm0.02$ for a large sample of optically selected quasars at $z\lesssim 3$. 

We note that at low-$z$, quasars can be divided into flat-spectrum, steep-spectrum, and peaked-spectrum sources (see e.g. review by \citealt{Urry1995PASP..107..803U}), which can also be expected to be the case at high-$z$. 
Our analysis cannot be used to probe spectral shape, however Fig.~\ref{fig:redshift_vs_alpha} illustrates the large intrinsic scatter in the spectral index of individual sources. 
An additional illustration of this potential variation in radio properties is the $z=6.82$ quasar, PSO J172.3556+18.7734, recently discovered by \citet[][outside the LoTSS-DR2 footprint]{Banados2021ApJ...909...80B}, which has a steep radio spectral index of $\alpha=$-1.31 between 1.4 and 5.0 GHz, but is not formally detected in LOFAR (with a $\sim0.3$ mJy/beam peak flux density). If the spectral slope of $\alpha=$-1.31 was continued, the predicted LOFAR flux would have been $\sim$20 mJy, therefore the non-detection in LOFAR clearly demonstrates a spectral turnover. 
The implications of the choice of spectral index on the radio-loud fraction are discussed in Sect.~\ref{sec:radio_loudness}. 

\begin{figure}
\centering
   \includegraphics[width=\columnwidth, trim={0.0cm 0cm 0cm 0.0cm}, clip]{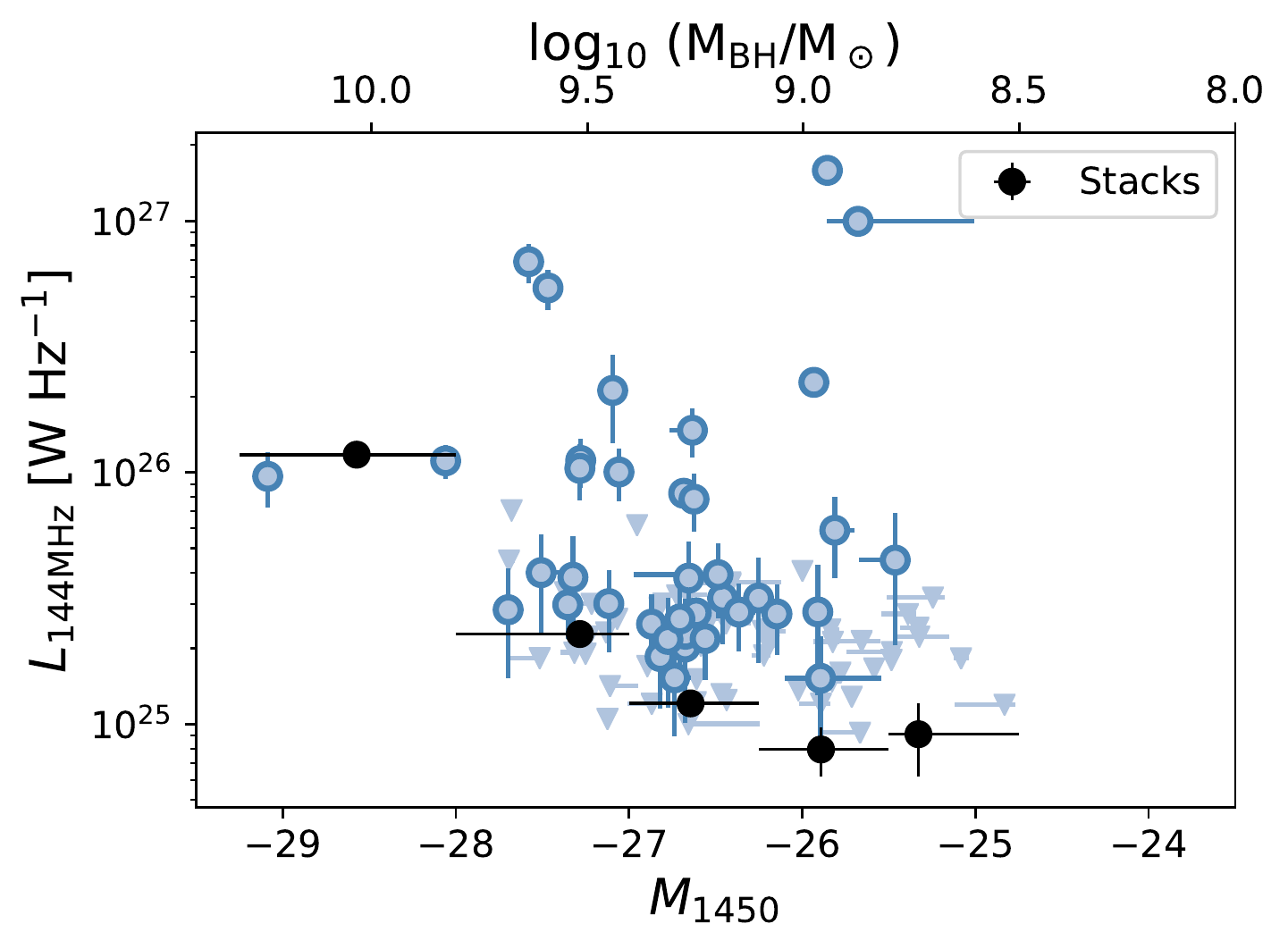}
     \caption{144 MHz radio luminosity as a function of absolute rest-frame UV magnitude at 1450 \r{A} determined by SED fitting with EAZY. The upper axis indicates the corresponding black hole masses based on the empirical relation given in Eq.~\ref{eq:bhmass}. The errors in $M_{1450}$ are given by the 16th and 84th percentile on the $M_{1450}$ distribution determined by randomly varying the measured fluxes within the error and refitting the SED.}
     \label{fig:UVlum_BHmass}
\end{figure}

Fig.~\ref{fig:radiolum_vs_z} shows the radio luminosity ($L_{144 \text{MHz}}$) versus spectroscopic redshift of the high-$z$ quasars detected by LOFAR. For non-detected quasars ($< 2\sigma$) within the LoTSS-DR2 area an upper limit is given. The spectral indices from Fig.~\ref{fig:redshift_vs_alpha} are used for the determination of the radio luminosities. For the majority of quasars the median spectral index of $-0.29 \pm 0.10$ is used, because these are not detected in FIRST. The shaded area shows the average LoTSS-DR2 5$\sigma$ depths as a function of redshift. This figure shows that the population of radio detected quasars is more numerous when going to lower radio luminosities, highlighting the importance of deep radio observations.
By stacking the radio continuum images at the optical positions for our sample we can measure the radio flux density (if the stacked image shows a detection) for the full sample of optically selected quasars within different redshift ranges. We divided the quasars in 4 redshift bins of $z=5$ to $7.0$ in steps of 0.5 and stacked the LOFAR cutouts in Fig.~\ref{fig:radiostacked_zbins}. In the first three redshift bins there are significant detections of $>5\sigma$, however in the final redshift bin $6.5<z<7.0$ the smaller number of sources results in only a measurement of $\sim$2$\sigma$. The radio luminosities derived from these stacks (again assuming $\alpha=-0.29$) are also given in Fig.~\ref{fig:radiolum_vs_z} and they are below the LoTSS-DR2 depth at all redshifts. To be able to probe the average radio power of this sample, the LOFAR survey would have to reach 5$\sigma$ sensitivities of $\sim$25 $\mu$Jy or better (a depth already reached in the designated LOFAR deep fields).

The previous result has shown we can detect quasars across a wide redshift range with LOFAR. This allows for studying how the radio emission correlates with the optical properties for this quasar sample. The radio luminosity as a function of $M_{1450}$, derived from SED fitting, is given in Fig.~\ref{fig:UVlum_BHmass}. Again, upper limits are given for the quasars not detected by LOFAR. On average the radio luminosity appears to be higher for quasars with brighter $M_{1450}$ magnitudes, as expected (see e.g. \citealt{Jiang2007ApJ...656..680J, Macfarlane2021}). From Fig.~\ref{fig:M1450_Z} and \ref{fig:radiolum_vs_z} there is no evidence for any strong trends of the absolute UV magnitude and radio luminosity of the quasar sample with redshift, therefore the full sample can be stacked to investigate other dependencies. Here, we stack the LOFAR cutouts in 4 bins of $M_{1450}$ and show the median radio luminosity and $M_{1450}$ in Fig.~\ref{fig:UVlum_BHmass}. In these values derived from the stacks an upward trend is confirmed with higher radio luminosities for brighter UV magnitudes. 

\begin{figure}
\centering
   \includegraphics[width=\columnwidth, trim={0.0cm 0cm 0cm 0.0cm}, clip]{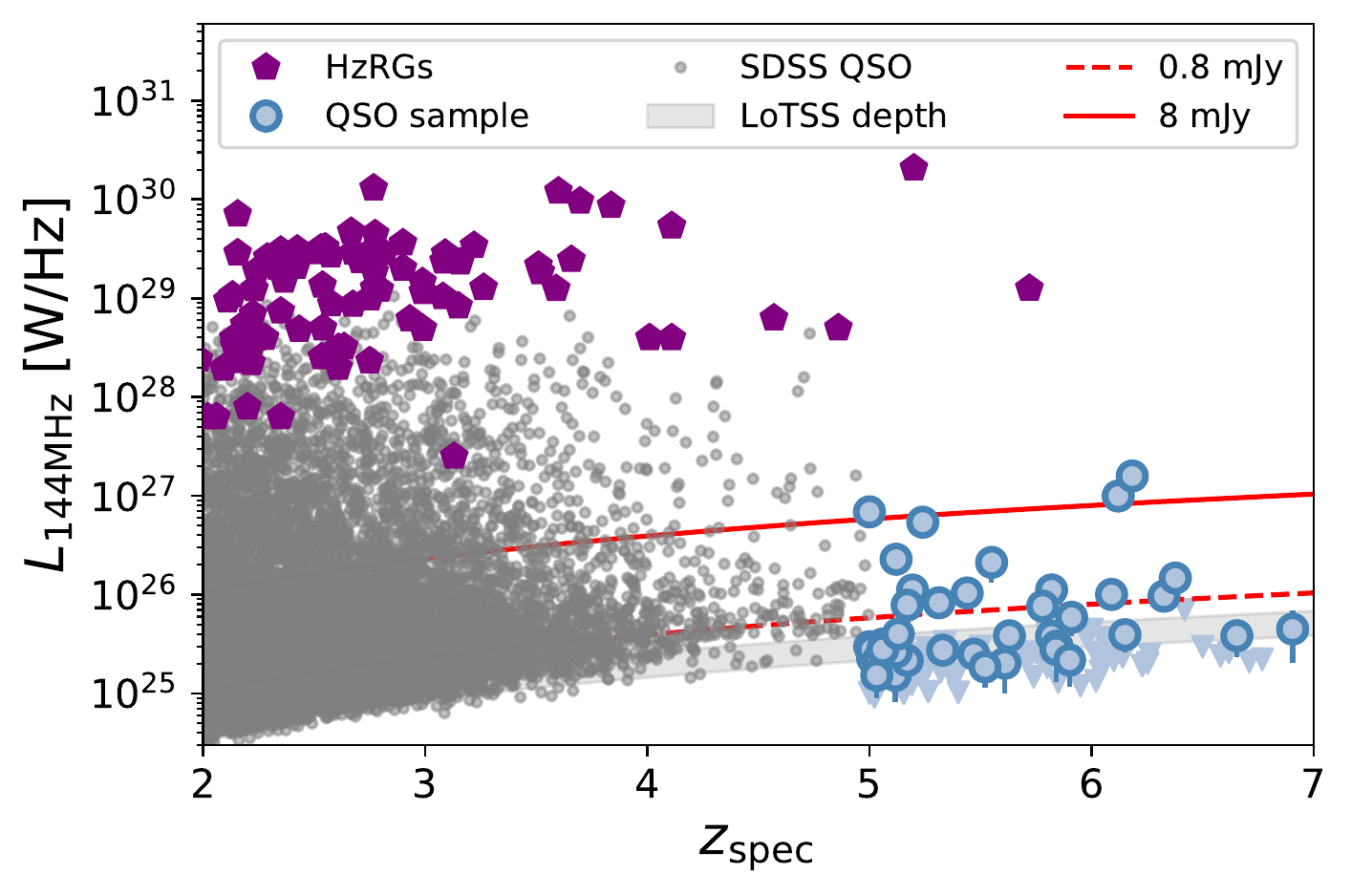}
     \caption{Radio luminosity of high-$z$ quasars compared to known radio galaxies (purple diamonds; \citealt{saxena2019MNRAS.489.5053S}) and SDSS DR16 quasars detected in LoTSS-DR2 between $2<z<5$ (grey circles) as a function of their redshift. The grey shaded area indicates the $5\sigma$ detection limit of LoTSS and the filled blue circles are quasar detections at a $>2\sigma$ level. Radio luminosities corresponding to 0.8 mJy and 8 mJy at 144 MHz are indicated with red dashed and solid lines respectively. The triangles indicate upper limits for quasars with $<2\sigma$ detection.}
     \label{fig:radiolum_vs_z_hzrgs}
\end{figure}

We can compare our quasar sample to the compilation of known high-redshift radio galaxies (HzRGs) presented by \cite{saxena2019MNRAS.489.5053S}. The majority of these HzRGs have been selected as steep spectrum sources (e.g. $\alpha^{150}_{1400} < -1.3$), and are therefore extremely bright at low-frequencies. It has been shown that these HzRGs are amongst the most massive galaxies in the Universe (see review of \citealt{Miley2008A&ARv..15...67M}). In Fig.~\ref{fig:radiolum_vs_z_hzrgs} the radio luminosities of the HzRGs are compared to our quasar sample and here it becomes apparent there is a large gap in radio luminosity between the VHzQ and HzRG samples, in the sense that the quasars have lower radio luminosities. However, this gap is closed at lower redshift ($z < 3$) where samples become significantly more complete. This highlights that this gap at higher redshift is likely due to observational limits and sample selection effects. The high-$z$ quasars are mostly selected by colour cuts from optical and near-infrared (NIR) surveys (e.g. \citealt{stoughton2002AJ....123..485S, Banados2016ApJS..227...11B, Miyazaki2018, wang2021ApJ...907L...1W}). Additionally, most HzRGs are selected due to being steep spectrum sources in the radio, which has been shown to be correlated with the redshift (e.g. \citealt{tielens1979A&AS...35..153T, Rottgering1994A&AS..108...79R, debreuck2000A&AS..143..303D}). However, not all HzRGs necessarily have a steep spectrum (e.g. \citealt{Yamashita2020AJ....160...60Y}) and in selecting high-$z$ quasars from optical surveys strict cuts are necessary to reduce the contaminating fraction of low-$z$ objects such as cool stellar dwarfs (e.g. \citealt{Fan_1999, Banados2016ApJS..227...11B, Matsuoka2018ApJS..237....5M}). A radio detection is therefore valuable information when selecting these candidate high-$z$ quasars, since the stellar dwarf contaminants of classical Lyman break and colour selection techniques typically do not show radio emission. 
Adding in a LOFAR detection criteria could allow a loosening of the colour cuts and give access to a larger potential sample of additional rare high-$z$ objects. Therefore, the combination of optical and radio detections may be highly useful for finding high-$z$ quasars and HzRGs that do not satisfy the traditional requirements. 

\section{Radio loudness distribution of high-$z$ quasars}
\label{sec:radio_loudness}

\begin{figure*}
\centering
   \includegraphics[width=\textwidth, trim={0.0cm 0cm 0cm 0.0cm}, clip]{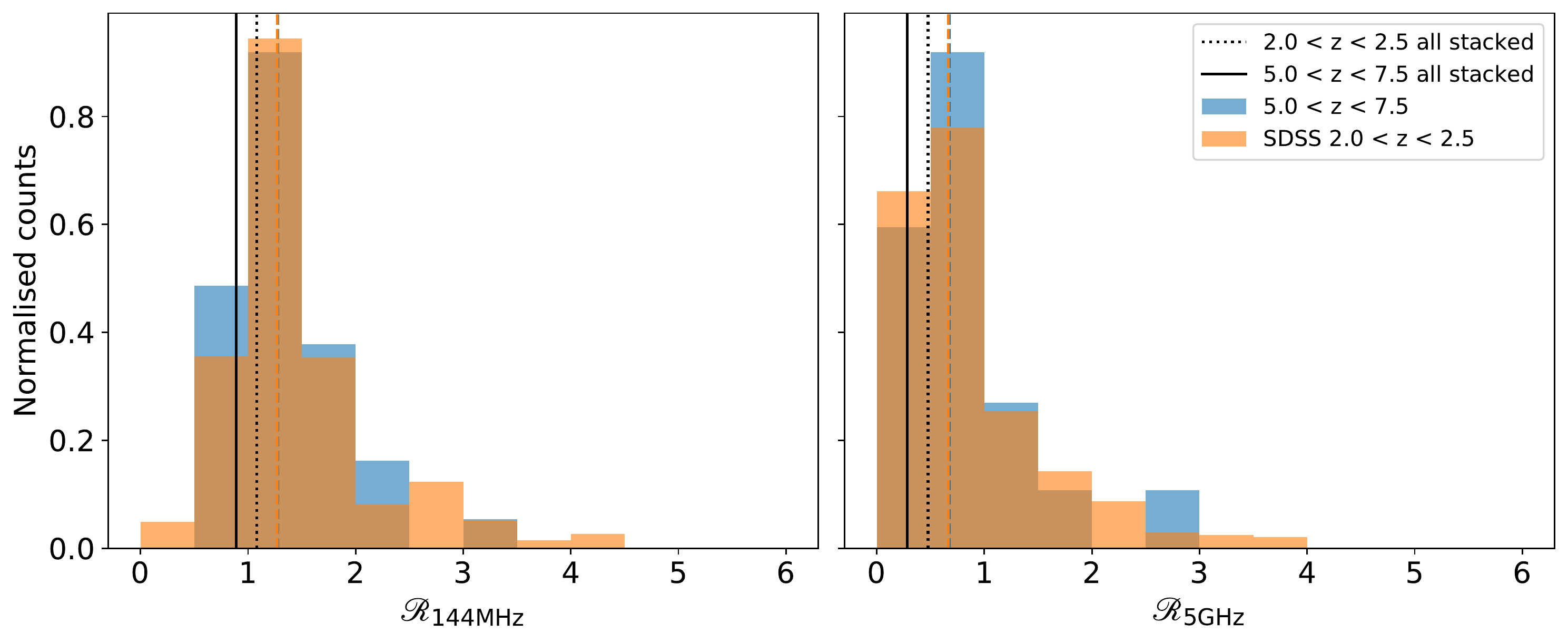}
     \caption{Normalised radio loudness distribution of only the LOFAR detected high-$z$ quasar sample (blue, 38 sources) and a magnitude matched sample of LOFAR detected SDSS quasars between $2.0<z<2.5$ (orange) using Eq.~\ref{eq:radioloudness150MHz} (left) and Eq.~\ref{eq:radioloudness5GHz} (right).The low-$z$ sample (3800 sources) is created by magnitude matching with the high-$z$ sample (see Sect.~\ref{sec:radio_loudness}), to ensure an equal underlying optical magnitude distribution. The coloured dashed lines indicate the median of the distributions. In the left panel these values are almost equal, causing the blue and orange dashed lines to almost overlap. The solid and dotted black line indicate the median radio loudness resulting from stacking all high-$z$ quasars and matched low-$z$ quasars, respectively.} 
     \label{fig:radioloudness}
\end{figure*}

Our understanding of the physics of SMBH accretion across the Universe's history can be enhanced by studying the evolution of jet powers. Additionally, acquiring the distribution of apparent radio flux densities at high-$z$ enables us to quantify the potential for 21-cm absorption studies. Even though the current sample of high-$z$ quasars is inhomogeneous due to the biased source selection, it does allow for making the first observational constraints on the radio loudness distribution at $z\sim6$, which can be used to test the redshift evolution. 

The radio loudness can be defined as the ratio of the radio and optical luminosity of a source. In the literature, a range of radio loudness definitions are used, with varying photometric bands and observed frame and rest-frame luminosities (see e.g. \citealt{Hao2014arXiv1408.1090H}). 
In this work the best accessible luminosity to use is $L_{1450}$ in the rest-frame UV. 
However, to be able to compare to the work of e.g. \cite{Banados2015ApJ...804..118B}, we also convert the $L_{144\text{MHz}}$ to rest-frame $L_{5\text{GHz}}$ using $\alpha = -0.29\pm0.10$ and use the $L_{4400}$ as rest-frame optical luminosity. The rest-frame 4400 $\AA$ is still well constrained in our SED fitting, because of the NIR and MIR observations in our quasar sample.
The two radio loudness parameters used in this work are
\begin{align}
\label{eq:radioloudness150MHz}
   \mathscr{R}_{144\text{MHz}} &= \log_{10} \Big( \frac{L_{144 \text{MHz}}}{L_{1450}} \Big) \\
   \mathscr{R}_{5\text{GHz}} &= \log_{10} \Big( \frac{L_{5 \text{GHz}}}{L_{4400}} \Big),
\label{eq:radioloudness5GHz}
\end{align}

\noindent with $L_{1450}$ and $L_{4400}$ the derived luminosity at 1450 $\AA$ and 4400 $\AA$ from SED fitting (see Sect.~\ref{subsec:sedfitting}). 

When determining the radio loudness distribution we include only the >2$\sigma$ detected sources by LOFAR and remove the quasar J1427+3312 selected by the FIRST survey, to ensure there is no preset selection bias towards radio loud quasars. The small size and inhomogeneous selection function of the high-$z$ sample means that detailed modelling of the distribution (as done by e.g. \citealt{Macfarlane2021}) is not warranted. However, we can check for consistency with distributions at lower redshift to test for any significant evolution. For a lower redshift sample we select known quasars from the Sloan Digital Sky Survey (SDSS) DR16 survey \citep{Lyke2020ApJS..250....8L} in the redshift range of $2.0<z<2.5$ with a LOFAR detection in the LoTSS-DR2 catalogue. This low-$z$ sample consists of 8627 quasars. The SEDs are again fitted using the same method in EAZY to obtain the rest frame $M_{1450}$ magnitudes. To be able to compare the radio loudness distributions, we randomly draw 100 quasars from the low-$z$ sample with $M_{1450}$ magnitudes within $\pm 0.05$ of each of the 38 quasars in the high-$z$ sample (thus building a total low-$z$ sample of 3800 sources). This ensures the UV magnitudes of the two distributions are the same and there is no bias towards fainter optical sources in the low-$z$ sample. Furthermore, we restrict the radio luminosity of low-$z$ quasars to the 2$\sigma$ limit at $z=6$ (of L$_{144\text{MHz}}$ = 10$^{25.2}$ W Hz$^{-1}$) to fix the low-$z$ sample to the same LOFAR detection limit. The normalised radio loudness distributions $\mathscr{R}_{144\text{MHz}}$ and $\mathscr{R}_{5\text{GHz}}$ of the high-$z$ and low-$z$ sample are shown in the left and right panels of Fig.~\ref{fig:radioloudness} respectively. The high-$z$ radio loudness distributions (blue) have a median and standard deviation found by bootstrapping of 1.27$\pm$0.51 and 0.68$\pm$0.58 for $\mathscr{R}_{144\text{MHz}}$ and $\mathscr{R}_{5\text{GHz}}$, respectively. The low-$z$ distributions (orange) are similar with a median and standard deviation of 1.26$\pm$0.68 and 0.65$\pm$0.70 for $\mathscr{R}_{144\text{MHz}}$ and $\mathscr{R}_{5\text{GHz}}$, respectively. Additionally, the black lines indicate the median radio loudness found by stacking all high-$z$ (solid line) and matched low-$z$ quasars (dotted line), which is as expected slightly lower than the median radio loudness of the radio detected sources. No significant deviation is found here between the low-$z$ and high-$z$ radio loudness distributions. However, note that the sample of high-$z$ quasars is small, consisting of only 38 sources. Therefore, only a considerable radio loudness evolution would be measurable. A noticeable difference in the radio loudness distributions is the absence of high-$z$ quasars in the tail of the distribution with $\mathscr{R} > 3.5$ (see Fig.~\ref{fig:radioloudness}), which is as expected due to the sample size. The sources in the tail of the $z=2$ distribution are by definition exceptionally radio bright and optically faint. A larger statistical sample of high-$z$ quasars is needed to investigate whether these sources exist at $z>5$. The lack of strong redshift dependence is in line with previous work (e.g. \citealt{gurkan2019A&A...622A..11G, Macfarlane2021}).

\begin{figure}
\centering
   \includegraphics[width=\columnwidth, trim={0.0cm 0cm 0cm 0.0cm}, clip]{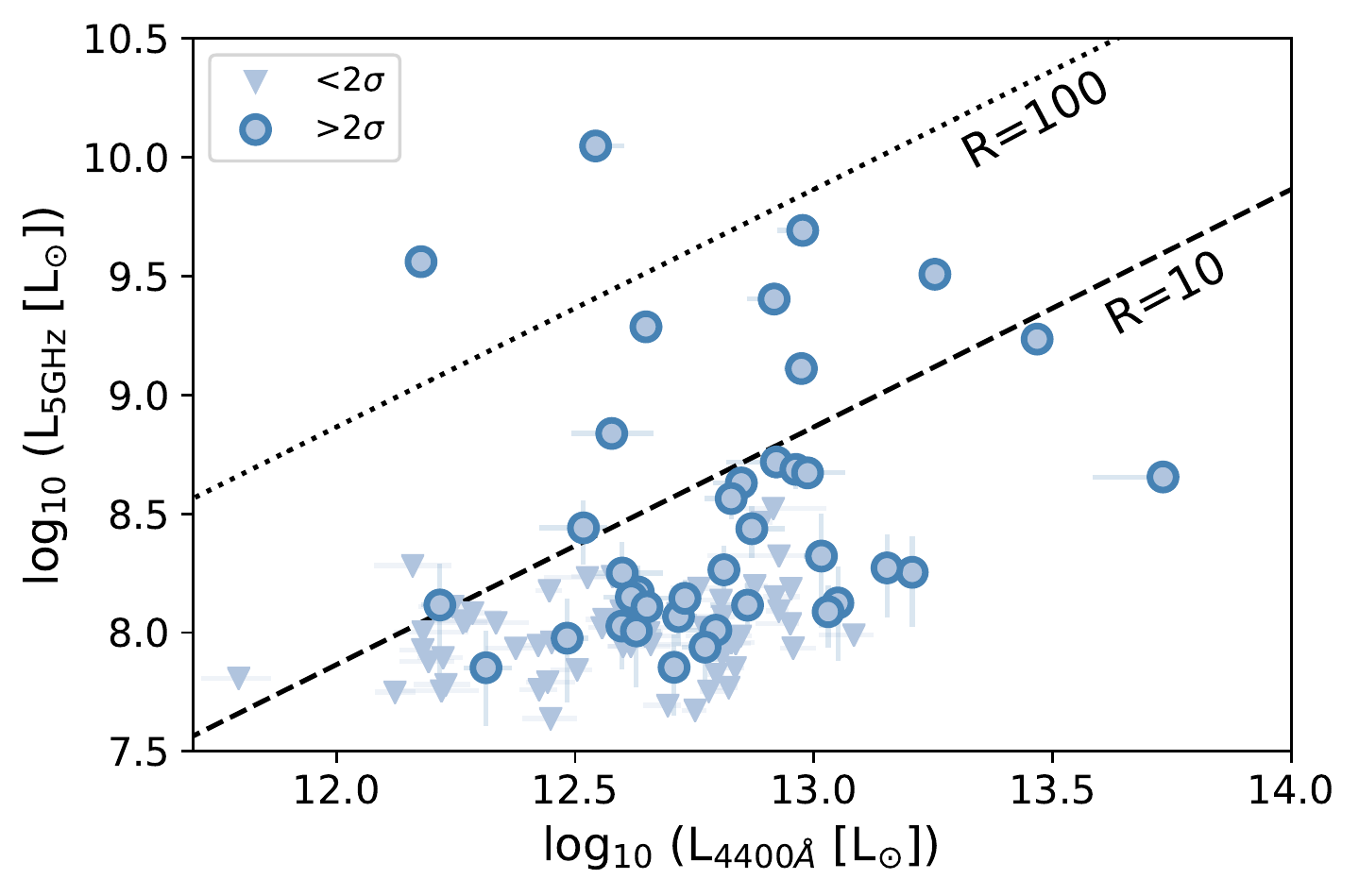}
     \caption{Radio luminosity at rest-frame 5 GHz versus optical luminosity at rest-frame 4400 $\AA$ for the high-$z$ quasar sample. The radio-loud criterion is defined as $R = f_{\nu,5\text{GHz}} / f_{\nu,4400\AA} > 10$ in line with other work, leading to a radio-loud fraction of this sample of $\sim$10\% if assuming the non-detected quasars are radio-quiet.}
     \label{fig:radioloudness_fraction}
\end{figure}

To determine the fraction of radio-loud quasars in our sample and to be able to compare our results to other work of e.g. \cite{Banados2015ApJ...804..118B}, we adopt the most widespread used radio-loud criterion, $R = f_{\nu,5\text{GHz}} / f_{\nu,4400\AA} > 10$, which should not be confused with $\mathscr{R}_{144\text{MHz}}$ and $\mathscr{R}_{5\text{GHz}}$ which are luminosity based (see Eq.~\ref{eq:radioloudness150MHz}). In the literature a quasar is defined as radio-loud if $R > 10$ and radio-quiet if $R < 10$. Following this dichotomy, the radio-loud fraction of our high-$z$ quasar sample (satisfying the optical detection criteria) is 10$\pm$2\%, assuming the non-detected quasars are radio-quiet (see Fig.~\ref{fig:radioloudness_fraction}), which is in line with \cite{Banados2015ApJ...804..118B}. 
Note that this fraction could be as high as 16\% if the three non-detected quasars with upper limits above $R > 10$ and the quasars on the border within the error are also radio-loud. It is also important to note that this fraction is highly dependent on the spectral index assumed. 
\cite{Banados2015ApJ...804..118B} found a radio-loud fraction of 8.1$^{+5.0}_{-3.2}$\% at $z\sim6$ (using 1.4 GHz observations), however they assume a steeper spectral index of $-0.75$. With this assumption we would find a radio-loud fraction of $\sim$6\%. This highlights that the spectral slopes assumed can have a significant effect on the radio-loud fraction. 

Multiple possible explanations are given in the literature to explain the observed radio loudness distribution, including dependence on properties such as BH spin \citep{Blandford1982MNRAS.199..883B}, BH mass \citep{Ho2002ApJ...564..120H}, and magnetic flux threading \citep{Tchekhovskoy2011MNRAS.418L..79T, Sikora2013ApJ...765...62S}. The magnetic flux threading mechanism is a promising hypothesis for the main driver of the radio emission of quasars, which predicts that quasars become radio-loud when a massive, cold accretion event follows an episode of hot accretion triggered by, for instance, a major merger \citep{SikoraBegelman2013ApJ...764L..24S}. This model could explain the wide range of radio powers observed and also the increase of the radio luminosity with accretion rate (optical luminosity) as we observe in Fig.~\ref{fig:UVlum_BHmass}. However, given the small sample size of the high-$z$ quasars in this work and heterogeneous selection function, the current data do not allow any discrimination between models. The LoTSS-DR2 dataset will allow extension of the work at low-$z$ of e.g. \cite{gurkan2019A&A...622A..11G} and \cite{Macfarlane2021} to much larger number statistics and enable a complete analysis from $z\sim0$ out to the extreme redshifts now probed in this work. 

\section{Predictions for high-$z$ quasar samples in current and future radio surveys}
\label{sec:predictions}

The surface density of known $z>5$ quasars $>2\sigma$ detected by LOFAR in the current LoTSS-DR2 area ($\sim$5720 deg$^2$) is 72 per 1000 deg$^{2}$. When the LoTSS survey encompasses the entire Northern sky, we can therefore expect to detect at least $\sim$150 known high-$z$ quasars with LOFAR, based solely on existing non-heterogeneous quasar samples. However, samples of known quasars in the EoR selected by optical and IR surveys are rapidly expanding. Additionally, spectroscopic follow-up of radio sources could reveal new high-$z$ populations (see Sect.~\ref{sec:properties_quasars_lotss}). Besides radio bright quasars, there also exist radio galaxies at $z>5$ as shown in Fig.~\ref{fig:radiolum_vs_z_hzrgs}. This also includes the furthest known radio galaxy at $z=5.7$ of \cite{Saxena2018MNRAS.480.2733S}, which is bright and should be easily detected by LOFAR, but lies outside of the LoTSS-DR2 footprint. 

A novel approach to finding new high redshift radio sources detected by LOFAR will be taken by the WEAVE-LOFAR (WL) survey \citep{smith2016sf2a.conf..271S}. The WEAVE instrument \citep{Dalton2014SPIE.9147E..0LD} is a multi-object spectrograph installed on the William Herschel Telescope and will have first light before the end of 2021. 
In collaboration with the LOFAR survey team it will obtain $\sim$1 million spectra of LOFAR-selected radio galaxies and will be a powerful tool for discovering new high-$z$ radio galaxies and quasars. The 8 mJy and 0.8 mJy limits shown in Fig.~\ref{fig:radiolum_vs_z_hzrgs} correspond to the planned flux density limits of the WL wide and mid tiers (including both prime and filler targets). One of the main aims of the wide tier is to obtain spectra for all $>8$ mJy LOFAR sources to discover new $z>6$ radio galaxies and quasars in an area of $\sim$ 6\,500 deg$^2$. Such sources can be used to perform 21-cm absorption spectroscopy with LOFAR and the Square Kilometer Array (SKA) in the future to get a direct measurement of the neutral intergalactic medium in the EoR. Furthermore, new photometric samples obtained by e.g. Euclid \citep{euclid2019A&A...631A..85E} will probe lower rest-frame UV magnitudes and also yield significant numbers of high-$z$ quasar detections with LOFAR at $z>7$, beyond the reach of WEAVE.

Because we do not find any evolution in the radio loudness between $z=2$ and $6$ (see Fig.~\ref{fig:radioloudness}), we can combine the currently known quasar radio properties at $z=2$ with the well constrained quasar UV luminosity function at $z=6$ (e.g. \citealt{Willott2007, McGreer2013, Matsuoka2018ApJ...869..150M}) to make a first order empirical prediction of the $z=6$ radio luminosity function (RLF), with the aim to obtain predictions for future high-$z$ quasar samples from LoTSS. We therefore combine the UV luminosity function of \cite{Matsuoka2018ApJ...869..150M} at $z=6$ with the radio luminosity distribution model of \cite{Macfarlane2021} at $z=2$. The \cite{Macfarlane2021} model assumes the radio emission from quasars can be modelled by two components: a star formation (SF) and a jet component. The SF component is parametrised by as a normal distribution, where the key parameters are the average SF ($\psi$) and standard deviation ($\sigma$) as a function of the quasar redshift and rest-frame magnitude. The jet component follows a power-law distribution with as key parameters the slope of the power-law, which was found to exhibit no significant evolution and hence was fixed, and a normalisation, which in this case was parametrised by $f$; the fraction of sources with jet powers above a threshold value. \cite{Macfarlane2021} constrained the evolution of these parameters as a function of redshift and rest-frame magnitude, which we interpolate/extrapolate to derive the radio power distribution as a function of those properties. We combine this \cite{Macfarlane2021} model with the $z=6$ UV luminosity function of \cite{Matsuoka2018ApJ...869..150M} to obtain a prediction of the $z=6$ RLF. For the rescaling of the radio luminosity to $z=6$ we draw random spectral indices from the initial bootstrapped spectral index distribution created in Sect.~\ref{sec:properties_quasars_lotss}.

\begin{figure}
\centering
   \includegraphics[width=\columnwidth, trim={0.0cm 0cm 0cm 0.0cm}, clip]{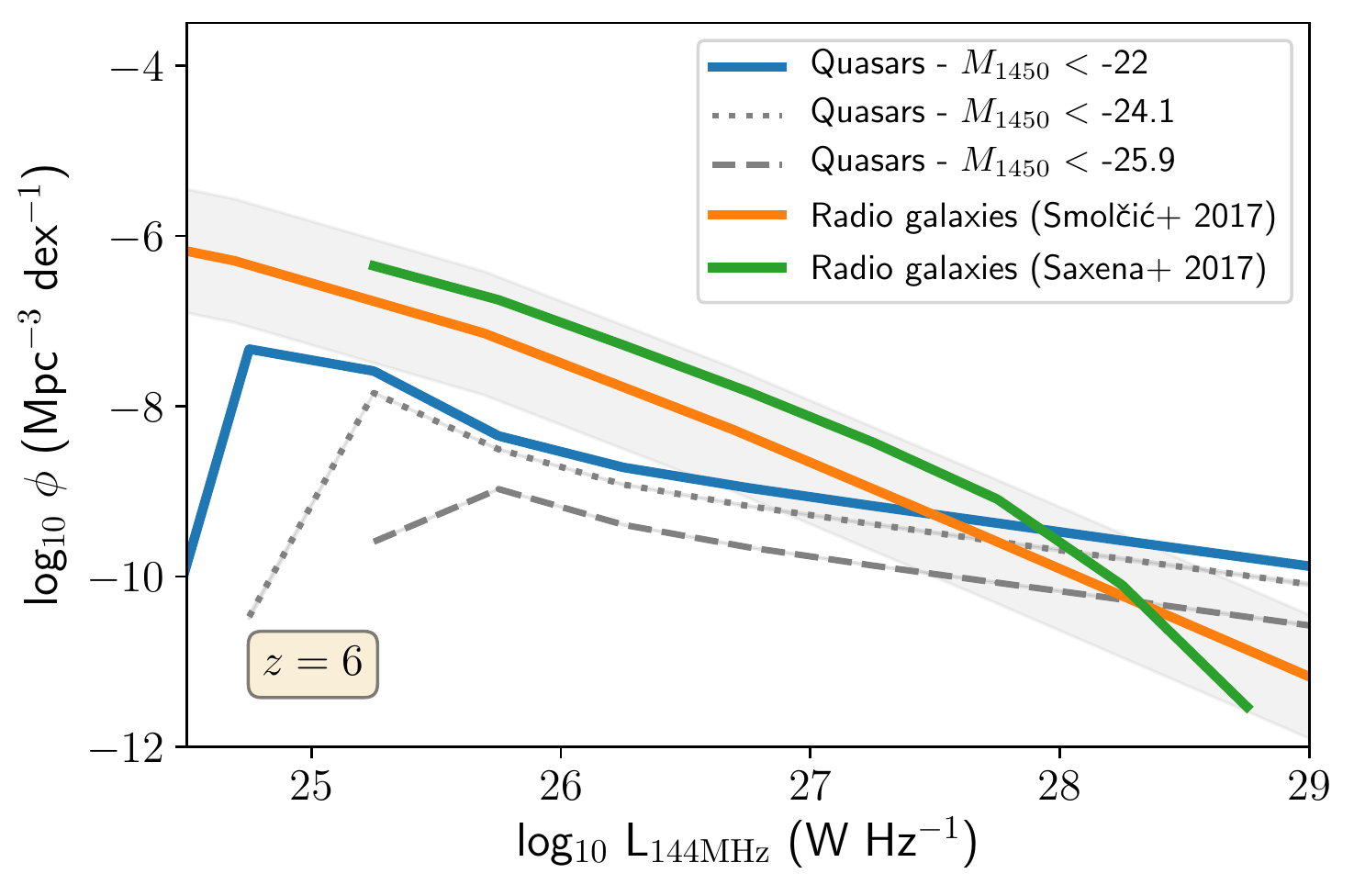}
     \caption{Radio luminosity function estimation of quasars (blue) based on the UV luminosity function of \cite{Matsuoka2018ApJ...869..150M} at $z=6$ and rescaled radio luminosity distributions of \cite{Macfarlane2021}. The RLF in blue is obtained using a $M_{1450}$ cut of -22, corresponding to future deep optical surveys like Euclid at $z=6$. The $M_{1450}$ limit of the Legacy and SDSS survey at $z=6$ are -24.1 (dotted line) and -25.9 (dashed line), respectively. The $z=6$ RLF is compared to the model of \cite{Saxena2017MNRAS.469.4083S} in green for a population of radio galaxies at $z=6$ and to the evolving radio-excess AGN pure density RLF fit from \cite{smolcic2017A&A...602A...6S} extrapolated to $z=6$ in orange. Since we extrapolate the fit of \cite{smolcic2017A&A...602A...6S} to higher redshift and luminosities, the bright end ($\gtrsim10^{26.5}$ W Hz$^{-1}$) is not well constrained.}
     \label{fig:rlf}
\end{figure}

The resulting quasar RLF is shown in Fig.~\ref{fig:rlf} for a UV magnitude range of -31.0 < $M_{1450}$ < -22.0, where the -22.0 limit at $z=6$ roughly corresponds to the detection limit of future surveys like the Euclid survey that extend much deeper than current all-sky surveys (see \citealt{euclid2019A&A...631A..85E}). For comparison we also show the RLFs with 5$\sigma$ limits of the Legacy and SDSS surveys, which correspond to -24.0 and -25.8 at $z=6$, respectively. These limits have been determined by converting 5$\sigma$ $z$-band limits to $M_{1450}$ limits using a $k$-correction obtained by taking the average $k$-correction for the quasar sample in LoTSS. The UV magnitude dependence of radio jet distributions in \cite{Macfarlane2021} is not constrained well beyond M$_{\text{UV}} < -26.0$, because of the small number of bright quasars in the area probed. Given the strong scaling between absolute magnitude and radio power observed at $z < 3$ in \cite{Macfarlane2021} (see \cite{Jiang2007ApJ...656..680J} for comparable effect in radio loud fraction), this may lead to increased uncertainty in the bright end of the RLF ($\gtrsim 10^{27}$ W Hz$^{-1}$). Also, we cannot rule out that the jet power distribution does not evolve at higher redshifts, which also contributes to the potential systematic uncertainty in these predictions. The $z=6$ RLFs for the different limits demonstrate that going from SDSS to Legacy depths results in the gain of most of the additional bright radio sources (i.e. a factor of $\sim$3 at 10$^{26}$ W Hz$^{-1}$), with much less gain when increasing the depth from Legacy to Euclid (a factor of $\sim$1.5 at 10$^{26}$ W Hz$^{-1}$). The majority of the most powerful radio-loud quasars at $z=6$ should therefore already be present in current optical datasets. 

The luminous radio AGN population \citep[$L_{144\text{MHz}}$ > 10$^{25}$ W Hz$^{-1}$;][]{Hardcastle2019A&A...622A..12H} is a combination of both quasars and radio galaxies, as also shown in Fig.~\ref{fig:radiolum_vs_z_hzrgs}. 
We therefore compare the quasar RLF in Fig.~\ref{fig:rlf} to the estimated radio galaxy RLF model of \cite{Saxena2017MNRAS.469.4083S}, as well as to the best-fit pure density RLF of the evolving radio-excess AGN RLF fit derived by \cite{smolcic2017A&A...602A...6S}. 
Although the $z=6$ RLF shown from \cite{smolcic2017A&A...602A...6S} represents an extrapolation from the dataset used to derive the fit ($z < 5.5$), it shows broad agreement with the RLF model of \cite{Saxena2017MNRAS.469.4083S}.

Additionally, Fig.~\ref{fig:rlf} illustrates that the radio AGN dominate the number counts for most radio luminosities. This is likely due to radio galaxies having long radio activity lifetimes, where the jets and lobes of the radio galaxies keep on growing. Conversely, quasars might have shorter duty cycles, where the radio activity quickly turns on and off, causing the lower number density. Also, some radio galaxies are likely to be obscured radio quasars, where the opening angle is thought to be luminosity dependent (see e.g. \citealt{Lawrence1991MNRAS.252..586L, Hao2005AJ....129.1795H, Treister2008ApJ...679..140T, Morabito2017MNRAS.469.1883M}), while there is also a population of radio galaxies that are jet-mode particularly at lower luminosities and are probably longer lived \citep{Heckman2014ARA&A..52..589H}.
The luminous radio galaxies of \cite{Saxena2017MNRAS.469.4083S} are expected to have a wide range of optical luminosities, with rest-frame UV/optical properties significantly below the detection limits of all-sky optical and infrared surveys. 
For example, the currently highest known redshift radio galaxy TGSS J1530+1049 at $z=5.7$ is non-detected in the optical/NIR ($J > 24.4$ and $K > 22.4$), but is very bright in the radio \citep[$L_{150\text{MHz}} = 10^{29.1} \text{W Hz}^{-1};$][]{Saxena2018MNRAS.480.2733S}. The \cite{Saxena2017MNRAS.469.4083S} prediction therefore models a complementary AGN population that is expected to be largely distinct from the samples probed by the known luminous quasar population.
Note that the quasar predictions here are based on the UV LFs, which in the case of \cite{Matsuoka2018ApJ...869..150M} are derived from colour-selected quasars. Expanding the colour selection could also result in larger quasar number counts in addition to the non-quasar radio galaxies.

To determine the expected number counts of quasars in the redshift range $5.5<z<6.5$ for different future radio surveys, we integrated the full optical depth (-31 < M$_{1450}$ < -22) $z=6$ quasar RLF (blue line in Fig.~\ref{fig:rlf}) down to the flux limits of the respective surveys given in Tab.~\ref{tab:expected_nums}. We can therefore expect to observe a few hundred high-$z$ $>8$ mJy sources in the WL survey, with many of these sources potentially missed by current optical quasar selections. 
In the full LoTSS survey we can eventually expect to detect a few thousand sources within the Northern hemisphere extragalactic sky ($\sim$10\,000 deg$^2$) if we assume the full depth reached by e.g. Euclid. However, if we are limited to sources with rest-frame UV magnitudes bright enough to be detected in SDSS or Legacy optical imaging, we would expect a number of $\sim$2400 and $\sim$810, respectively.
Finally, we also estimate the number counts for the Evolutionary Map of the Universe (EMU; \citealt{Norris2011PASA...28..215N}), a deep wide-area radio survey planned for the Australian Square Kilometre Array Pathfinder (ASKAP) telescope, which will survey the Southern Sky and part of the Northern Sky up to +30$^{\circ}$ declination at 1.3 GHz with RMS of $\sim$10 $\mu$Jy. 
Our current estimates from the LoTSS survey, suggest that $\sim$30,000 high-$z$ radio sources will be detected in the EMU survey at $5.5<z<6.5$. However, note that all above quasar number count predictions might be optimistic, since this calculation does not take into account radio losses due to CMB photons at high-$z$ and the possible quasar evolution beyond $z>6$. In addition, the calculation for the EMU survey is also strongly dependent on the assumed spectral index, and a steeper spectral index would greatly reduce this number. Here we assumed that the median spectral index value calculated from LoTSS-DR2 of $-0.29\pm$0.10 also applies down to the flux limit of EMU. For steep-spectrum radio sources ($\alpha$<-1.0) LOFAR is almost equally sensitive at 144 MHz as EMU at 1.3 GHz. 

For the radio fluxes predicted to be needed for EoR experiments (for example > 10 mJy or >10$^{27}$ W Hz$^{-1}$ at $z\sim6$), the radio galaxy population is expected to dominate the total number counts based on the predictions of \citet[][see Fig.~\ref{fig:rlf}]{Saxena2017MNRAS.469.4083S}. These radio galaxies generally have steeper spectral indices and are therefore brighter at low frequencies (see e.g. \citealt{gurkan2019A&A...622A..11G}). The combination of LOFAR and WL offers a unique way of finding these sources and the quasar luminosity function derived here can therefore be regarded as a lower limit on the expected number counts from the radio AGN population. 

\begin{table}
\caption{Estimation of expected number of $5.5<z<6.5$ quasars in the WEAVE-LOFAR, LoTSS, and EMU survey obtained from the full optical depth RLF (by integrating the blue line in Fig.~\ref{fig:rlf}) and corresponding survey parameters. The predicted ranges are based on the uncertainties of the UV LF and spectral index. In the case of the WEAVE-LOFAR number counts, the sources above $0.8$ and $8$ mJy will be targeted in the WL-wide and WL-mid tier respectively, but this does not guarantee that these will be robustly identified high-$z$ sources due to their optical properties.} 
\label{tab:expected_nums}      
\centering
\resizebox{\columnwidth}{!}{
\begin{tabular}{c c c c}  
\hline\hline       
Survey & Sky cov. & Flux lim. & Expected number counts\\ 
& (deg$^2$) & (mJy) &  Full depth\\
\hline 
WL & 6500 & 8 & 610 - 760 \\
 & 650 & 0.8 & 180 - 200\\
LoTSS & 10,000 & 0.5* & 3600 - 4000 \\ 
EMU & 15,000 & 0.1** &  30,000 - 31,000 \\ 
\hline \hline
\end{tabular}}
* 5$\sigma$ flux limit of LoTSS \\
** The 5$\sigma$ flux limit converted from 1.3 GHz to 144 MHz, using $\alpha = -0.29$
\end{table}

\section{Summary}
\label{sec:conclusions}

To summarise, the depth and coverage of the new LoTSS-DR2 data allows for studying the low-frequency properties of high-$z$ quasars for the first time. We detect 36\% of the known $z>5$ quasars with $>2\sigma$ in the LoTSS-DR2 area with LOFAR, which demonstrates that radio observations are a valuable addition to existing selection criteria of high-$z$ quasar candidates for follow-up spectroscopic observations by decreasing contamination and changing selection biases.

The median spectral index obtained by stacking 93 sources in LoTSS and FIRST is $-0.29^{+0.10}_{-0.09}$, which is in line with expectations from observations of quasars at $z < 3$. Stacking in redshift bins shows that LOFAR surveys would need to reach a 5$\sigma$ flux density limit of $\sim$25 $\mu$Jy to be able to probe the average radio power of this quasar sample. We do not find a strong evolution in radio loudness distribution when comparing our high-$z$ sample to a magnitude-matched lower-$z$ sample of $2.0<z<2.5$ SDSS DR16 quasars, but do note a larger statistical sample is needed to be able to detect weak evolution. We find a fraction of radio-loud high-$z$ quasars of 10$\pm$2\%, under the assumption that high-$z$ known quasars undetected in the LoTSS-DR2 at the 2sigma level are radio-quiet. This fraction is dependent on the spectral index assumed, which will be lower for steeper spectral indices.

We estimate a RLF at $z=6$, which predicts that a few hundred quasars with $>8$ mJy in the range of $5.5<z<6.5$ will be observed with optical spectroscopy in the upcoming WEAVE-LOFAR (WL) survey. The WL survey will take optical spectra of $\sim$1 million radio sources and will have great potential for finding high-$z$ quasars and radio galaxies in an unexplored parameter space, since no biases are introduced by optical selection techniques and radio steep spectrum criteria. For the future EMU survey, we predict to find thousands of high-$z$ sources at $>0.1$ mJy from the RLF, which will enable the study of radio properties of large statistical samples of high-$z$ quasars.

\begin{acknowledgements}
{KJD acknowledges funding from the European Union’s Horizon 2020 research and innovation programme under the Marie Sk\l{}odowska-Curie grant agreement No. 892117 (HIZRAD).
HJAR acknowledges support from the ERC Advanced Investigator programme NewClusters 321271. MJH acknowledges support from the UK Science and Technology Facilities Council (ST/R000905/1).
PNB is grateful for support from the UK STFC via grants ST/R000972/1 and ST/V000594/1. AD acknowledges support by the BMBF Verbundforschung under the grant 05A20STA. WLW acknowledges support from the CAS-NWO programme for radio astronomy with project number 629.001.024, which is financed by the Netherlands Organisation for Scientific Research (NWO).

LOFAR is the Low Frequency Array designed and constructed by ASTRON. It has observing, data processing, and data storage facilities in several countries, which are owned by various parties (each with their own funding sources), and which are collectively operated by the ILT foundation under a joint scientific policy. The ILT resources have benefited from the following recent major funding sources: CNRS-INSU, Observatoire de Paris and Université d'Orléans, France; BMBF, MIWF-NRW, MPG, Germany; Science Foundation Ireland (SFI), Department of Business, Enterprise and Innovation (DBEI), Ireland; NWO, The Netherlands; The Science and Technology Facilities Council, UK; Ministry of Science and Higher Education, Poland; The Istituto Nazionale di Astrofisica (INAF), Italy.

This research made use of the Dutch national e-infrastructure with support of the SURF Cooperative (e-infra 180169) and the LOFAR e-infra group. The J\"ulich LOFAR Long Term Archive and the German LOFAR network are both coordinated and operated by the J\"ulich Supercomputing Centre (JSC), and computing resources on the supercomputer JUWELS at JSC were provided by the Gauss Centre for Supercomputing e.V. (grant CHTB00) through the John von Neumann Institute for Computing (NIC).

This research made use of the University of Hertfordshire high-performance computing facility and the LOFAR-UK computing facility located at the University of Hertfordshire and supported by STFC [ST/P000096/1], and of the Italian LOFAR IT computing infrastructure supported and operated by INAF, and by the Physics Department of Turin university (under an agreement with Consorzio Interuniversitario per la Fisica Spaziale) at the C3S Supercomputing Centre, Italy.

This paper is based (in part) on data obtained with the International LOFAR Telescope (ILT) under project codes LC0 015, LC2 024, LC2 038, LC3 008, LC4 008, LC4 034 and LT10 01. LOFAR \citep{vanHaarlem2013A&A...556A...2V} is the Low Frequency Array designed and constructed by ASTRON. It has observing, data processing, and data storage facilities in several countries, which are owned by various parties (each with their own funding sources), and which are collectively operated by the ILT foundation under a joint scientific policy. The ILT resources have benefited from the following recent major funding sources: CNRS-INSU, Observatoire de Paris and Universit\'e d'Orl\'eans, France; BMBF, MIWF-NRW, MPG, Germany; Science Foundation Ireland (SFI), Department of Business, Enterprise and Innovation (DBEI), Ireland; NWO, The Netherlands; The Science and Technology Facilities Council, UK; Ministry of Science and Higher Education, Poland.}
	
\end{acknowledgements}

\bibliographystyle{aa}
\bibliography{bibliography.bib}

\end{document}